\documentclass[12pt,a4paper,epsf]{article}
\usepackage{graphics}
\usepackage{amssymb,amsmath}
\usepackage[dvips]{lscape,graphicx}
\usepackage{cite}
\usepackage{longtable}

\newcommand{\ct}{\cite}
\newcommand{\lb}{\label}

\newcommand{\bc}{\begin{center}}
\newcommand{\ec}{\end{center}}
\newcommand{\bd}{\begin{displaymath}}
\newcommand{\ed}{\end{displaymath}}
\newcommand{\be}{\begin{equation}}
\newcommand{\ee}{\end{equation}}
\newcommand{\ba}{\begin{array}}
\newcommand{\ea}{\end{array}}
\newcommand{\bea}{\begin{eqnarray}}
\newcommand{\eea}{\end{eqnarray}}
\newcommand{\bt}{\begin{tabular}}
\newcommand{\et}{\end{tabular}}
\newcommand{\un}{\underline}
\newcommand{\ov}{\overline}
\newcommand{\bp}{\begin{picture}}
\newcommand{\ep}{\end{picture}}
\newcommand{\bfi}{\begin{figure}}
\newcommand{\efi}{\end{figure}}

\def\fun#1#2{\lower3.6pt\vbox{\baselineskip0pt\lineskip.9pt
\ialign{$\mathsurround=0pt#1\hfil##\hfil$\crcr#2\crcr\sim\crcr}}}

\begin{document}



\vspace{1cm}

\title{\LARGE \bf New Bound States\\ of Top and Beauty Quarks \\at the Tevatron
and LHC}
\author{\Large C.R.~Das ${}^{1}$
\footnote{crdas@cftp.ist.utl.pt}\,\,, C.D.~Froggatt ${}^{2}$
\footnote{c.froggatt@physics.gla.ac.uk}\,\,, L.V.~Laperashvili
${}^{3}$ \footnote{laper@itep.ru}\,\, and
H.B.~Nielsen ${}^{4}$ \footnote{hbech@nbi.dk}\\[5mm]
\itshape{\large ${}^{1}$ Centre for Theoretical Particle
Physics,}\\ \itshape{\large Technical University of Lisbon,
Lisbon, Portugal}\\ \itshape{\large ${}^{2}$ Department of Physics
and Astronomy,}\\ \itshape{\large Glasgow University, Glasgow,
Scotland}\\ \itshape{\large ${}^{3}$ Institute of Theoretical and
Experimental Physics,}\\ \itshape{\large Moscow, Russia}\\
\itshape{\large ${}^{4}$ The Niels Bohr Institute, Copenhagen,
Denmark }}

\date{}

\maketitle

\thispagestyle{empty}

\begin{abstract}
The present paper is based on the assumption that heavy quarks
bound states exist in the Standard Model (SM). Considering  New
Bound States (NBS) of top-anti-top quarks (named T-balls) we have
shown that: 1) there exists the scalar $1S$--bound state of
$6t+6\bar t$; 2) the forces which bind the top-quarks are very
strong and almost completely compensate the mass of the twelve
top-anti-top-quarks in the scalar NBS; 3) such strong forces are
produced by the Higgs-top-quarks interaction with a large value of
the top-quark Yukawa coupling constant $g_t\simeq 1$. Theory also
predicts the existence of the NBS $6t + 5\bar t$, which is a color
triplet and a fermion similar to the $t'$-quark of the fourth
generation. We have also considered the ``b-quark-replaced'' NBS.
We have estimated the masses of the lightest fermionic NBS:
$M_{NBS}\gtrsim 300$ GeV, and discussed the larger masses of
T-balls. Searching for heavy quarks bound states at the Tevatron
and LHC is discussed.
\end{abstract}

\thispagestyle{empty}

\clearpage\newpage

\section{Introduction}

Although the Standard Model (SM) was confirmed by all experiments
of the world accelerators, the mechanism of the Electroweak (EW)
symmetry breaking (EWSB) has not yet been tested. According to the
SM, the Higgs boson is responsible for generating the masses of
fermions due to the  Higgs mechanism. However, the mass of the
Higgs boson is not predicted by theory.

Direct searches in the previous experiments (mainly at LEP2\,
\ct{1}) set a lowest limit for the Higgs boson mass $M_H$:
\be M_H \gtrsim 114.4 \,\,\, GeV\,\,\, at \,\,\, 95\% \,\,\, CL.
\lb{1a} \ee
The recent Tevatron result \ct{2} is:
\be 115 \lesssim M_H \lesssim 160\,\,\, GeV. \lb{2a} \ee
We hope that LHC will provide a solution of main puzzles of EWSB.

The Higgs boson couples more strongly to the heavy top quarks than
to the light ones. As a result, the Higgs exchanges between top
quarks produce new type of bound states
\ct{3,4,5,6,7,8,9,10,11,12,13,14,15}.

The present paper is devoted to the properties of the new bound
states (NBS): estimates of their masses and observation at modern
colliders (Tevatron, LHC, etc.). The predictions of
Refs.~\ct{3,4,5,6,7,8,9,10,11,12} are:

$\bullet$ There exists a scalar $1S$--bound state of $6t + 6\bar
t$. The forces which bind these top-quarks are so strong that
almost completely compensate the mass of the 12 top-quarks forming
this bound state.

$\bullet$ There exists a new bound state $6t + 5\bar t,$ which is
a fermion similar to the quark of the fourth generation having
quantum numbers of top quark.

$\bullet$ Theory also predicts the existence of new bound states
with b-quark replaced the t-quark: for example, NBS $ n_b b + (6t
+ 6\bar t - n_b t)$, etc., where $n_b=1,...6.$

A new (earlier unknown) bound state $6t+6\bar t,$ which is a color
singlet (that is, `white' state), was first suggested by Froggatt
and Nielsen in Ref.~\ct{4}. Now all these NBS are named T-balls,
or T-fireballs.

\section{Higgs and gluon interactions of quarks}

If the Higgs particle exists, then between quarks $qq$, quarks and
anti-quarks $q\bar q$, and also between anti-quarks $\bar q\bar q$
there exist virtual exchanges by Higgs bosons (see Fig.~1),
leading only to the attractive forces.

 \bfi \centering
\includegraphics[height=80mm,keepaspectratio=true,angle=0]{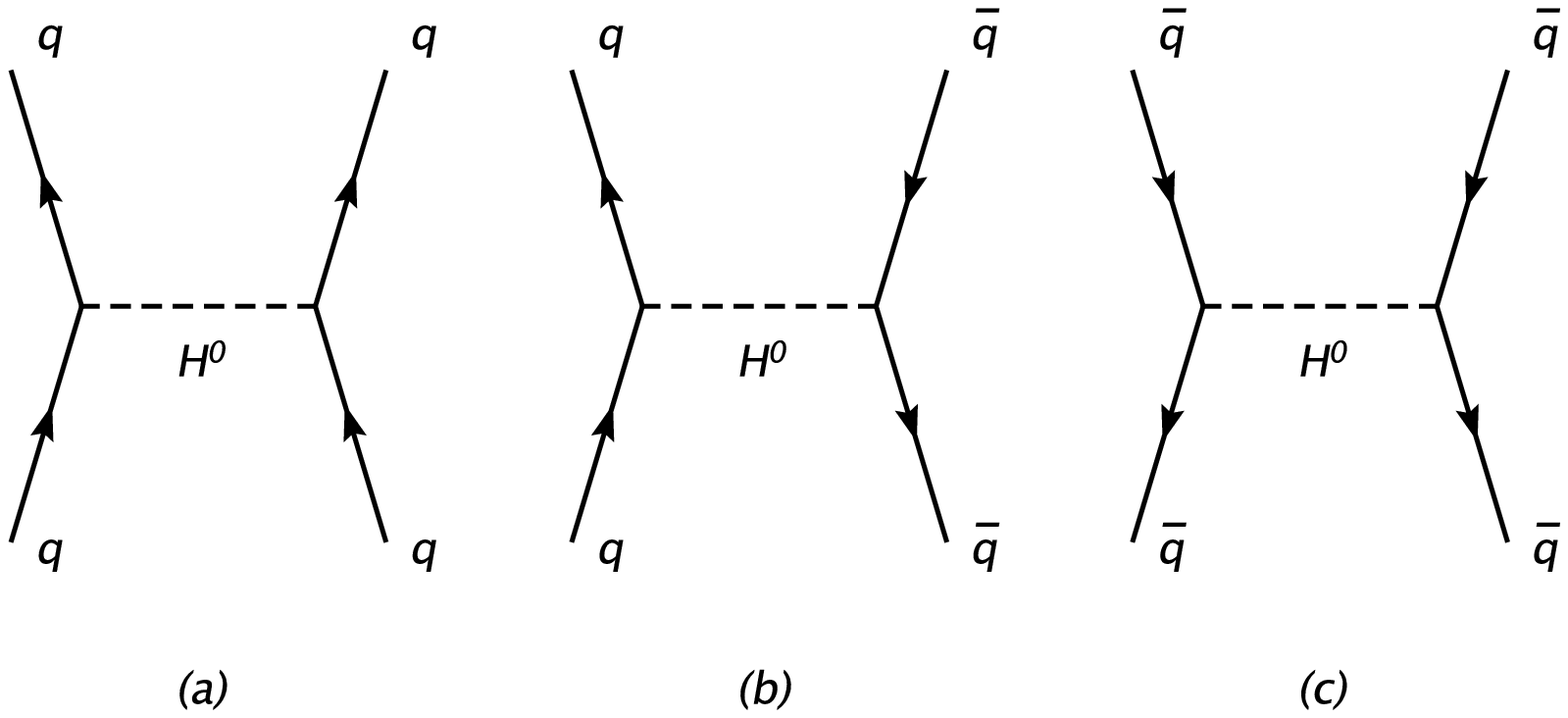}\caption{}\efi

It is well-known that the bound state $t\bar t$ -- so called
toponium -- is obliged to the gluon virtual exchanges of Fig.~2.
Among a considerable quantity of articles devoted to the toponium,
we distinguish the following backward papers
\ct{16,17,18,19,20,21}.

\bfi \centering
\includegraphics[height=80mm,keepaspectratio=true,angle=0]{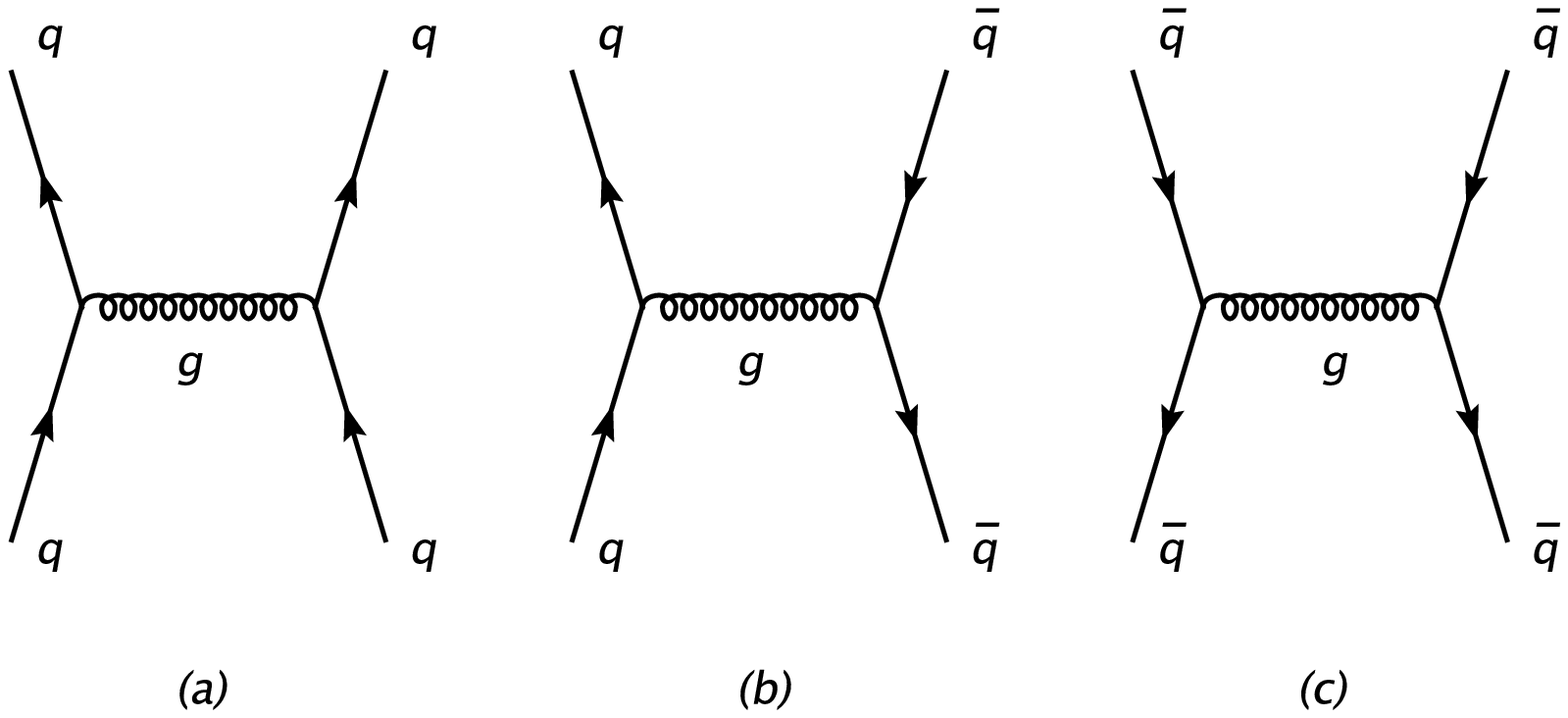}\caption{}\efi

In the case of the toponium the contributions of the Higgs scalar
particles are essential, but less than gluon interactions.
Toponium is very unstable due to the decay of the top quark
itself. However, putting more and more top and anti-top quarks
together in the lowest energy bound states, we notice that the
attractive Higgs forces continue to increase. Simultaneously gluon
(attractive and repulsive) forces first begin to compensate
themselves, but then begin to decrease relatively to the Higgs
effect with growth of the number of top-anti-top constituents in
the NBS.

The maximum of the binding energy value corresponds to the
$1S$-wave state of the NBS $6t + 6\bar t$ . The explanation is
simple: top-quark has two spin states and three states of colors:
$2\times 3=6$ degrees of freedom.  This means that, according to
the Pauli principle, only 6 pairs of $t\bar t$ can simultaneously
exist in the `white' $1S$-wave state. If we try to add more $t\bar
t$-pairs , then some of them will turn out to the $2S$-wave state,
and the NBS binding energy will decrease at least 4 times. For
P-,D-, etc. wave states the NBS binding energy decreases more and
more.

\section{ T-ball mass estimate.}

The kinetic energy term of the Higgs field and the top-quark
Yukawa interaction are given by the following Lagrangian density:
\be
          L = \frac 12 D_{\mu}\Phi_H D^{\mu}\Phi_H + \frac{g_t}{\sqrt
          2}\ov{\psi_{tL}}\psi_{tR}\Phi_H  + h.c.,   \lb{1} \ee
where $\Phi_H$ and $\psi_t$ are the Higgs and top-quark fields,
respectively, and $g_t$ is the Yukawa coupling constant of their
interaction.

The VEV of the Higgs field in the EW-vacuum is:
\be v=<|\Phi_H|>=246\,\, {\rm{GeV}}.\lb{2} \ee
According to the Salam-Weinberg theory the top-quark mass $M_t$
and the Higgs mass $M_H$ are given by the following relations:
\be  M_t = \frac {g_t}{\sqrt 2}v \quad {\rm{and}} \quad M_H^2
=\lambda v^2, \lb{3} \ee
where $\lambda$ is the Higgs self interaction coupling constant.

According to the Ref.~\ct{22}, we have \be M_t\approx
172.6\,\,{\rm{ GeV}},\lb{4} \ee
and \be g_t \approx 0.93. \lb{5} \ee
Let us imagine now that the NBS is a bubble in the EW-vacuum and
contains $N_{const.}$ top-like constituents. It is known that
insight the bubble (bag) the Higgs field can modify its VEV.
Implications related with this phenomenon have been discussed in
Refs.~\ct{5,21,23,24,25,26,27}. Then insight T-balls the VEV of
the Higgs field is smaller than $v$:
\be v_0 = <|\Phi_h|>,\quad {\rm{where}}\quad \frac{v_0}{v} <
1,\lb{6} \ee
and the effective masses insight the bubble (bag) are smaller than
the corresponding experimental masses:
\be m_{t,h} = \frac{v_0}{v}M_{t,H}.\lb{7} \ee
In this case the attraction between the two top (or anti-top)
quarks is presented by the Yukawa type of potential:
\be  V(r) = - \frac{g_t^2/2}{4\pi r}\exp(-m_hr). \lb{8} \ee
Assuming that the radius $R_0$ of the bubble is small:
\be m_hR_0 << 1,\lb{9} \ee
we obtain the Coulomb-like potential: \be V(r) \simeq -
\frac{g_t^2/2}{4\pi r}. \lb{10} \ee
The attraction between any pairs $\Large \bf tt,\,\,t\bar
t,\,\,\bar t\bar t$ is described by the same potential (\ref{10}).

By analogy with Bohr Hydrogen-atom-like model, the binding energy
of a single top-quark relatively to the nucleus containing
$Z=N_{const.} - 1$ top-quarks have been estimated in
Refs.~\ct{4,5,6}. The total potential energy for the NBS with
$N_{const.}= 12$ is:
\be  V_{tot}(r) = - 11\frac{g_t^2/2}{4\pi r}. \lb{11} \ee
Here we would like to comment that the value of the mass $m_h$,
which belongs to the Higgs field insight the NBS $6t + 6\bar t$,
can just coincide with estimates given by Refs.~\ct{13,14,15}. The
results: $\rm max (m_h)=29$ Gev and $\rm max (m_h)=49$ Gev
correspond to Ref.~\ct{13} and Ref.~\ct{15}, respectively.

Considering a set of Feynman diagrams (the Bethe-Salpeter
equation) and including the contributions of all (s-,t- and u-)
channels for the Higgs and gluon exchange forces (see Ref.~[6]),
we obtain the following Taylor expansion:
\be
    M_T^2 = (N_{const.}M_t)^2\times $$
    $$       \left\{1 -
     2(N_{\mbox{const.}}-1){\left(\frac{
N_{const.}}{12}\right)}^2\left(\frac{g_t^2 + \frac 16
g_s^2}{\pi}\right)^2 +....\right\}. \lb{12} \ee
Here the QCD coupling constant $g_s$ is given by its fine
structure constant value at the EW-scale \ct{22}:
\be
       \alpha_s(M_Z) = g_s^2(M_Z)/4\pi \approx 0.118. \lb{14} \ee
Now the value of the total binding energy for arbitrary
$N_{const.}$ is equal to:
 \be
    E_T = N_{const.}(N_{const.} - 1){\left(\frac{ N_{const.}}{12}\right)}^2
    {\left(\frac{g_t^2 + \frac 16 g_s^2}{\pi}\right)}^2
    m_t. \lb{15} \ee
The mass of T-ball containing $N_{\mbox{const.}}$ top or anti-top
quarks is:
 \be  M_T = N_{\mbox{const.}}m_t - E_T.\lb{16} \ee

Approximately this dependence is described by the following
expression:
 \be   M_T = N_{\mbox{const.}}m_t\left\{1 -
(N_{\mbox{const.}}-1){\left(\frac{
N_{const.}}{12}\right)}^2\left(\frac{g_t^2 + \frac 16
g_s^2}{\pi}\right)^2\right\}.   \lb{17} \ee

Below we shall use the following notations: $T_s$-ball is a scalar
NBS $6t + 6\bar t$, having the spin $S=0$, and $T_f$-ball presents
the NBS $6t + 5\bar t$, which is a fermion: $\ov {T_f} = 5t +
6\bar t$.

Let us consider now the condition:
\be  \frac{11}{\pi^2}\cdot (g_t^2 + \frac 16g_s^2)^2 = 1. \lb{18}
\ee
In this case  the binding energy $ E_T$ compensates the NBS mass $
12 m_t$ so strongly that the mass of the scalar $ T_s$-ball
becomes zero:
 \be  M_{T_s} = 11m_t\left\{1- \frac{11}{\pi^2}\cdot (g_t^2 + \frac 16g_s^2)^2\right\}=
 0. \lb{19} \ee
It is necessary to emphasize that the experimental values given by
(\ref{5}) and (\ref{14}) \ct{22}:
\be g_t^2\simeq 0.86 \quad {\rm {and}} \quad g_s^2\simeq 1.48
\lb{20} \ee
are just very close to this limit.

Fig.~3 shows the dependence of T-ball masses on the number of NBS
constituents $N_{\mbox{const.}}$. In the case when $ M_{T_s} = 0$,
we have:
\be M_T = N_{\mbox{const.}}m_t\left\{1 -
\frac{(N_{\mbox{const.}}-1)}{11}\frac{N_{const.}^2}{12^2}\right\}
\lb{21} \ee

\bfi \centering
\includegraphics[height=100mm,keepaspectratio=true,angle=0]{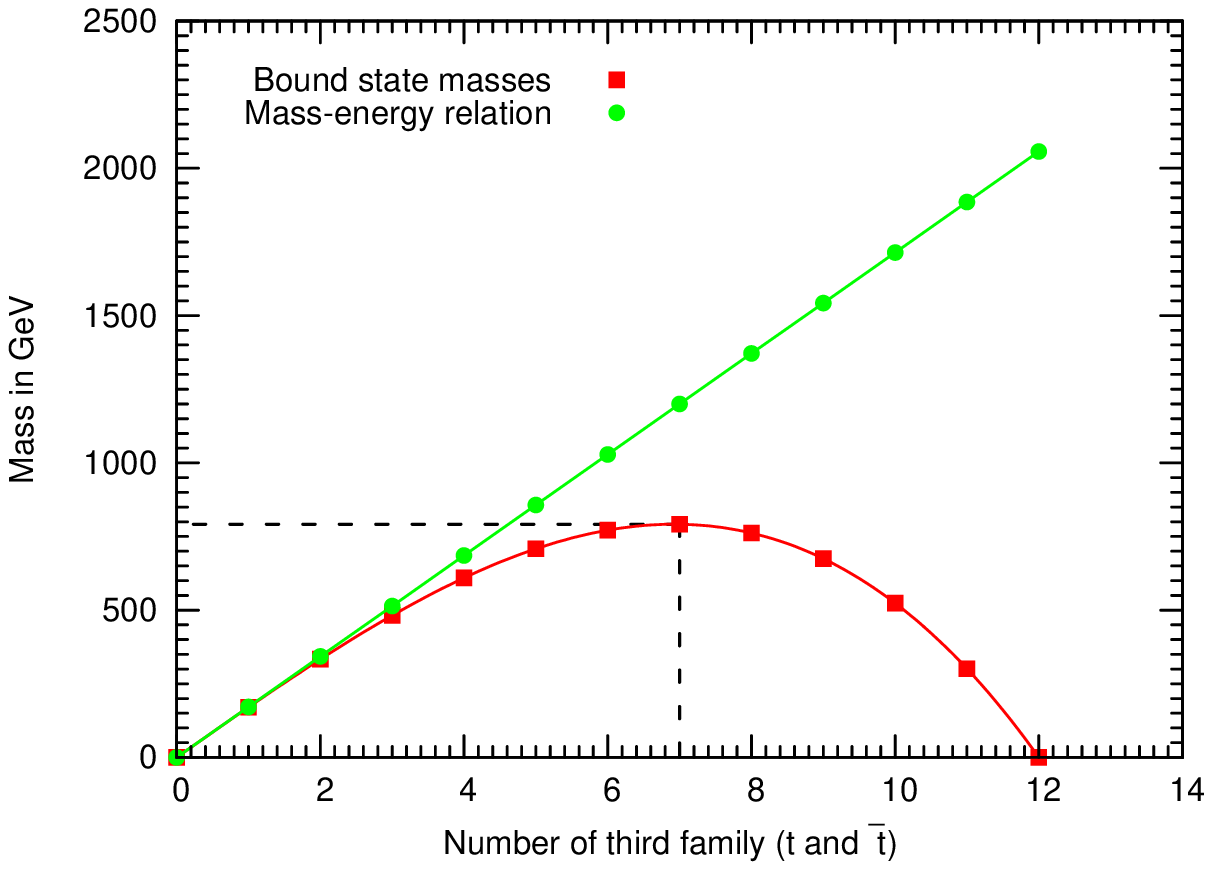}
\caption{T-ball mass depending on the number $N_{\mbox{const.}}$
of the NBS constituents. } \lb{4f}\efi

We easily see that the light scalar Higgs bosons with mass $ m_h <
M_H$ can bind the 12 top-like quarks so strongly that the mass
$M_{T_s}$ becomes almost zero, and even tachyonic: $ M_{T_s}^2 <
0.$ In the last case we obtain the Bose-Einstein condensate of
T-balls -- a new vacuum at the EW-scale \ct{11,12}. Previously the
condensation of $t\bar t$, arising from four-fermion interaction
models (\ct{28,29,30}, etc.), was reviewed in Ref.~\ct{31}. We
have suggested a new type of condensation of top-quarks via
T-balls, what is very important for the solution of the hierarchy
problem in the SM \ct{9,10}.

\subsection{$T_f$-ball mass estimate}

As we have discussed above, the Higgs interaction of the eleven
top-anti-top quarks ($N_{\mbox{const.}}=11$) creates a $T_f$-ball
-- a new fermionic bound state $6t + 5\bar t$, which is similar to
the $t'$-quark of the fourth generation. The estimate of the mass
of $T_f$-ball $6t + 5\bar t$ by Eq.~(\ref{21}) gives :
\be  M_{T_f}\approx 11m_t\cdot 0.236\,\,\gtrsim\,\, 300\,\, GeV.
\lb{22} \ee
We hope that the forthcoming numerical calculations of the T-ball
masses by Monte-Carlo simulations on lattice will give us more
exact answers.

\section{New ``b-replaced'' bound states}

Constructing T-balls from $t$- and $\bar t$-quarks, we also can
take into account considerable contributions of left b-quarks
insight NBS \ct{3,6,11}.

If we had no $b\bar b$-pairs in T-balls, then there would be an
essential superposition of different states of the weak isospin.
The presence of b-quarks in the NBS leads to the dominance of the
isospin singlets of EW-interactions only. Now such a
``b-replaced'' scalar NBS would be stable. We predict the
following scalar ``b-replaced'' NBS:
\be T_s(b-replaced) = b + 5t + 6\bar t, \lb{B1} \ee
\be T_s(\bar b-replaced) = 6t + \bar b + 5\bar t. \lb{B2} \ee
In general case we can construct the following scalar
``b-replaced'' T-balls:
\be T_s(nb-replaced) = n_b b + (6t + 6\bar t - n_b t), \lb{B3} \ee
and
\be T_s(n\bar b-replaced, ) = n_{\bar b}\bar b + (6t + 6\bar t -
n_{\bar b}\bar t). \lb{B4} \ee
Of course, we also can construct the fermionic ``b-replaced''
NBS:
\be T_f(b-replaced)= b + 5t + 5\bar t, \lb{B5} \ee
and
\be \ov{T_f}(\bar b-replaced)= 5t + 5\bar t + \bar b. \lb{B6} \ee
In general case we obtain:
 \be T_f(nb-replaced) =
n_b b + (6t + 5\bar t - n_b t), \lb{B7} \ee
and
\be \ov{T_f}(n\bar b-replaced) = n_{\bar b}\bar b + (5t + 6\bar t
- n_{\bar b}\bar t). \lb{B8} \ee
We have $n_b, n_{\bar b}=1,...6$ in Eqs.~(\ref{B3})-(\ref{B8}).

 There is a simple way to estimate the mass of the
``b-replaced'' T-ball with one t-quark replaced by a b-quark. It
is well-known that b-quark does not interact significantly with
NBS. Thus, we can add a b-quark (or anti-b-quark) to the NBS
having eleven constituents without essential changing its energy,
or mass. Then the b-replaced scalar NBS $T_s(b-replaced)$, or
$T_s(\bar b-replaced)$, given by Eqs.~(\ref{B1}) and (\ref{B2}),
respectively, will have a mass $\,\,\backsimeq 300$ GeV.

As to the NBS $T_f(b-replaced) = 5t + b + 5\bar t$ and $
T_f(b-replaced, b\bar b) = 5t + b + n_b b\bar b + 5\bar t$, they
will have a mass very close to the NBS with ten constituents, e.g.
$M_{T_f} \backsimeq 500$ GeV (see Fig.~3).

We also can consider more heavy T-balls with $M_T > 500$ GeV, but
they will have very small cross-sections of their production.

\section{Can we observe T-balls at LHC or Tevatron?}

If our NBS are strongly bound states with small radius, they can
be observed at colliders (Tevatron, LHC, etc.) in the following
processes:

1) First of all, in the possible H-decay process:
 \be H\to 2T_s, \lb{23} \ee
 if $M_{T_s} < \frac 12 M_H.$ Using limits given by Tevatron experiments \ct{2}:
 $115 \lesssim M_H \lesssim 160\,\,\, GeV,$ we obtain the requirement for the Higgs
decay mechanism:
\be M_{T_s} \lesssim 80 \,\, GeV.  \lb{24} \ee
Here we have argued that T-balls can explain why it is difficult
to observe the Higgs boson H at colliders: T-balls can strongly
enlarge the decay width of the Higgs particle.

2) If $M_{T_s} > \frac 12 M_H$, then the first decay (\ref{23}) is
absent in Nature, and the $T_s$-balls fly away, forming jets which
produce hadrons with a high multiplicity:
\be T_s \to JETS. \lb{24} \ee
3) Second, we can observe at Tevatron all processes given by
Fig.~4 with the replacement $$t\bar t \to t'\bar
t',\,\,T_f\ov{T_f}.$$ In the most optimistic cases the NBS $ 6t +
5\bar t$ (fermionic fireball) plays a role of the fundamental
quark of the fourth generation, say, with the mass $M_{T_f}\gtrsim
300$ GeV, given by our preliminary estimate. We expect that the
Tevatron-LHC experiments should find either a fourth family
t'-quark, or the fermionic NBS $T_f $, or both of them.

\bfi \centering
\includegraphics[height=120mm,keepaspectratio=true,angle=0]{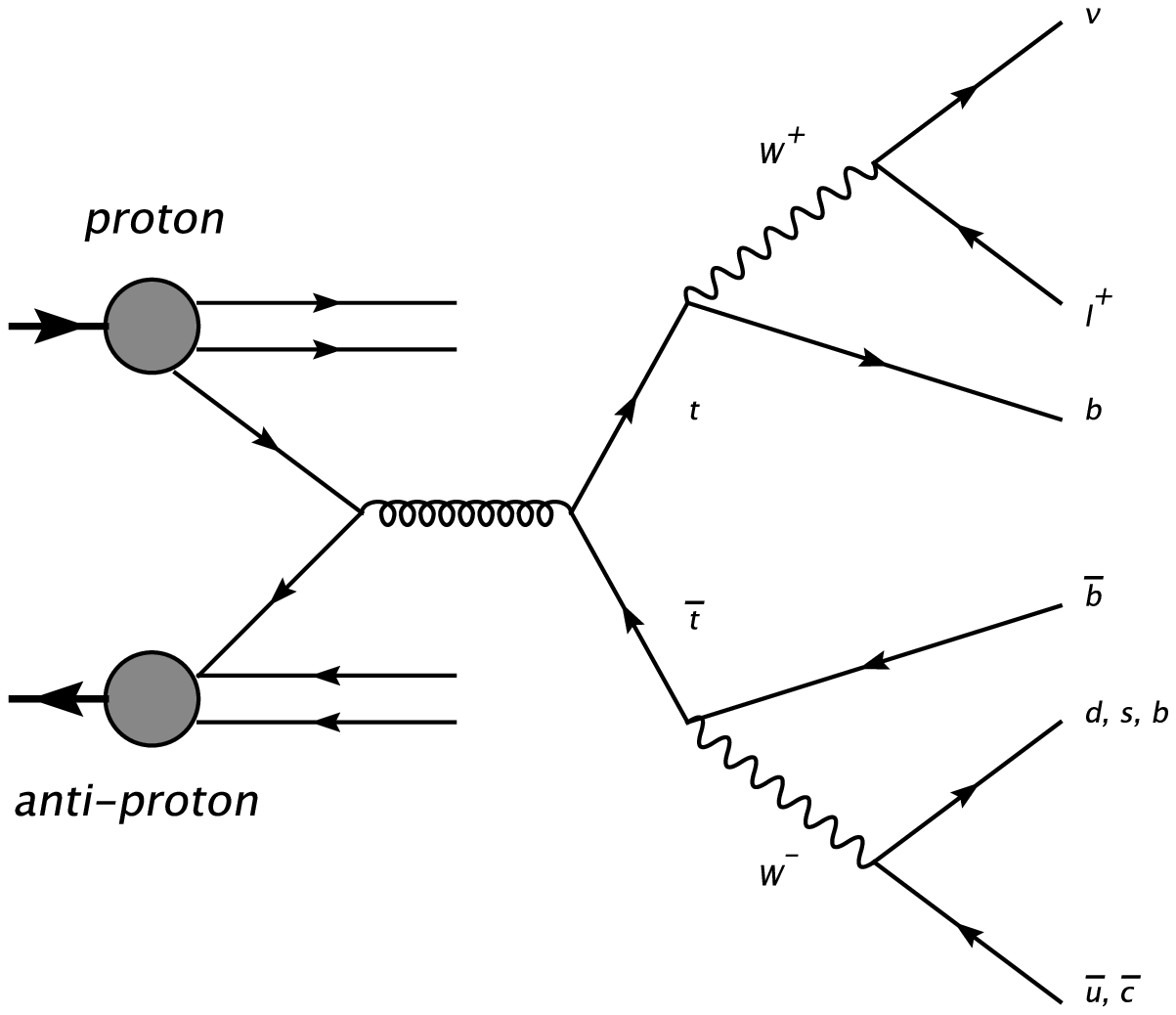}
\caption{A typical process observed at the Tevatron in $p\bar p$
collisions.}\efi

The scalar NBS $T_s$ cannot be produced simply in a pair by a
gluon vertex, because it is a color singlet $\un{1}$. But a pair
$T_f\ov{T_f}$ can be produced by a gluon, because $T_f$ is a color
triplet $\un{3}$.

At LHC the pairs of $T_s$-balls, or $T_f$-balls might be produced
in $pp$ collisions via the two gluon diagram with strong vertices
shown in Fig.~5 \ct{3}.

\bfi \centering
\includegraphics[height=70mm,keepaspectratio=true,angle=0]{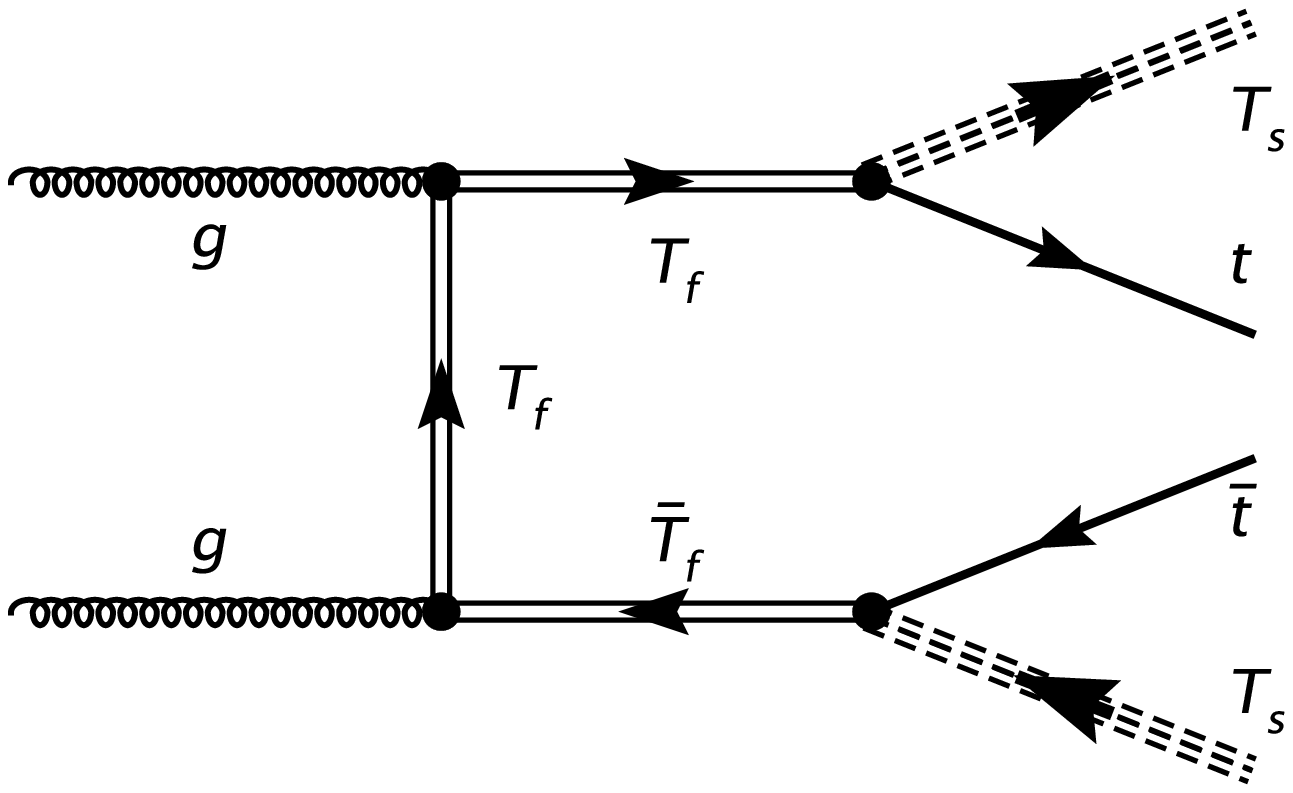}
\caption{Two gluon production of $T_s$-balls} \efi

\section{CDF II Detector experiment at the Tevatron}

Recent experiments with CDF II Detector of the Tevatron \ct
{32,33} searching for heavy top-like quarks in $p\bar
p$-collisions with $ \sqrt s \backsimeq 1.96$ TeV do not exclude
the existence of T-balls with masses  $\gtrsim
300\,\,{\mbox{GeV}}$ up to 500 GeV.

Here we can assume that the very strange events observed at the
Tevatron as a fourth family $t'$, which decays into a $ W$-boson
and a presumed quark-jet, might find another explanation in our
model: maybe it is a decay of T-balls into a $W$-boson and a gluon
jet.

Tevatron experiments exclude a fourth-generation t' quark with a
mass below 300 GeV (see Refs.~\ct{32,33}).  Assuming that fourth
generation $t'$-quarks does not exist in Nature, but only the
pairs of fermionic NBS $T_f$ are produced at the Tevatron, we can
give an explanation of the observed cross-sections shown in
Fig.~6. The curve for the cross-section
\be \sigma(p\bar p \to t'\bar t')\simeq 0.1\,\, pb \lb{24} \ee
can correspond to the production of pairs of fermionic $T_f$-balls
with mass $M_{T_f} \gtrsim 300$ GeV.

\bfi \centering
\includegraphics[height=120mm,keepaspectratio=true,angle=0]{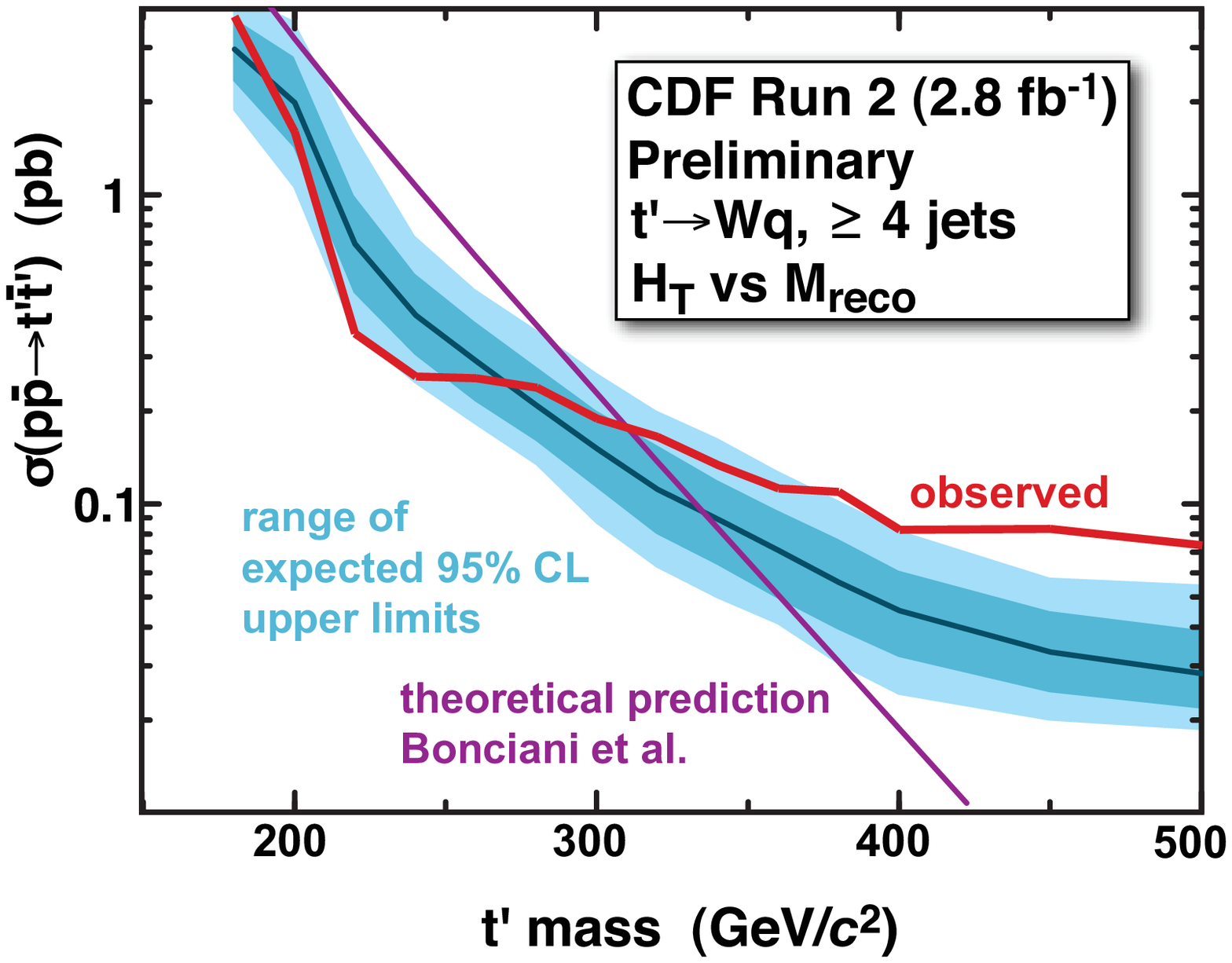}
\caption{Tevatron CDF-experiment given by Refs.~\ct{32,33}: upper
limit, at 95\% CL, a fourth-generation t' quark with a mass below
300 GeV is excluded. Blue line presents a theoretical curve for
the fourth-generation quarks cross-section.}\efi

\section{Estimate of the NBS form-factors in the Tevatron CDF-experiment}

Assuming that the fourth-generation $t'$-quarks does not exist in
Nature, but only the fermionic $T_f$-balls with mass $M_T
> 300$ GeV are produced at the Tevatron in
the CDF-experiment \ct{32,33}, we can imagine the existence of
form-factors of the NBS $T_f$, which determine the cross-section
of the production of the fermionic T-balls (see Fig.~6):
\be \sigma(p\bar p\to T_f\ov{T_f})=F^2(M_T)\sigma_{theor}(M_T).
\lb{1ff} \ee
Here $\sigma(p\bar p\to T_f\ov{T_f})$ is given by the observed red
line curve of Fig.~6 and $\sigma_{theor}(M_T)$ is given by the
theoretical (blue) curve obtained by Bonciani et al. \ct{34,35}
for the point-like particle $t'$. Our numerical calculations of
the form-factor shown in Fig.~7 gives the results in the region of
$M_T$ from 311 GeV (where $F(M_T)=1$) up to 500 GeV. We conclude
that for $M_T=500$ GeV the form-factor is large enough:
\be
       F(M_T)\approx 7.6. \lb{2ff}
\ee

\bfi \centering
\includegraphics[height=100mm,keepaspectratio=true,angle=0]{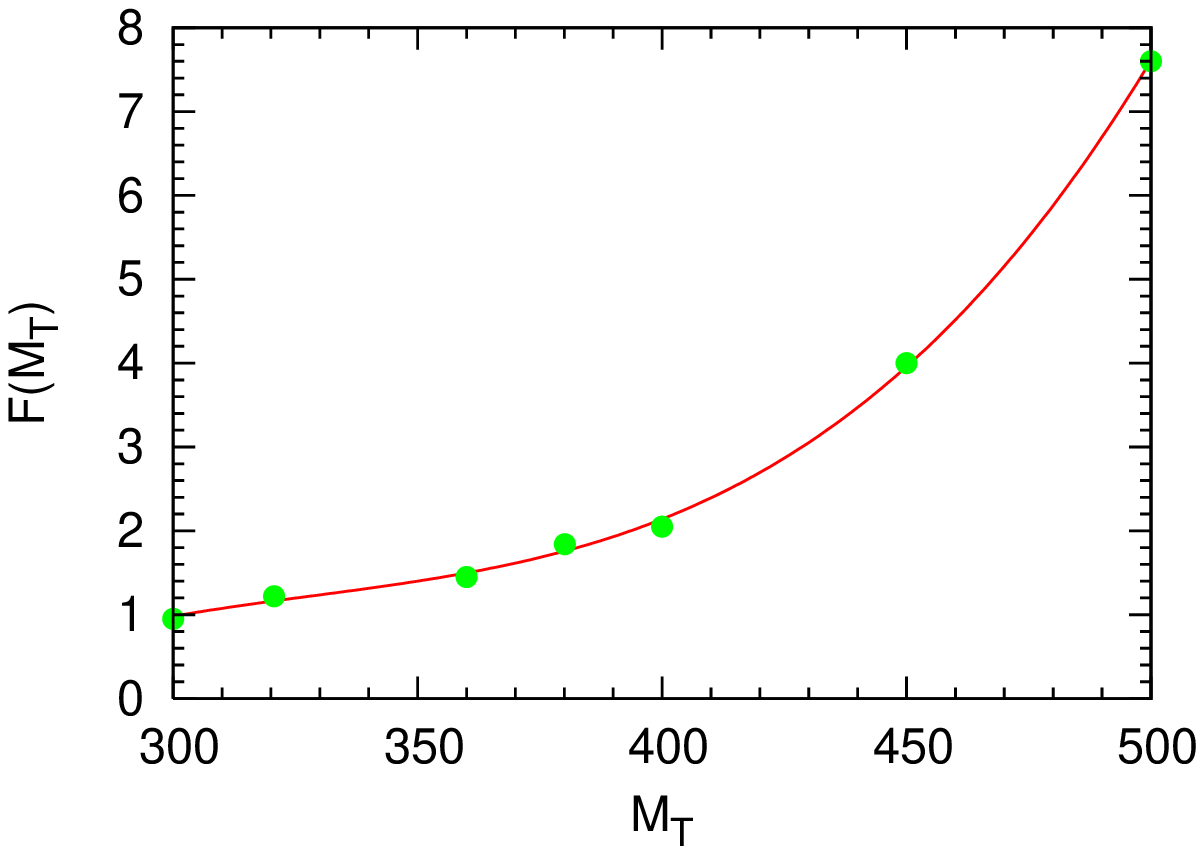}
\caption{The form-factor $F(M_T)$ of the fermionic new bound state
$T_f$ obtained from Tevatron CDF-experiment \ct{32,33} in absence
of the four generation.}\efi

\section{Conclusions}

At present, a lot of physicists, theorists and experimentalists,
are looking forward to the New Physics. However, it is quite
possible that LHC will discover only the Salam-Weinberg Higgs
boson and nothing more. Nevertheless, T-balls considered in the
present paper could exist in the framework of the SM.

The present investigation is based on the assumption that there
exist in Nature new bound states of top-like quarks, so called
T-balls, or T-fireballs. First, we predict that there exists
$1S$--bound state of $6t+6\bar t$. The forces in the NBS can bind
top-like quarks so strongly that they can almost completely
compensate the mass of the 12 top-quarks in the scalar bound
state. Such strong forces are produced by interactions of
top-quarks via the virtual exchanges of the scalar Higgs bosons,
when the top-quark Yukawa coupling constant is large: $g_t\simeq
1$.

Present theory also predicts the existence of the new bound state
$6t + 5\bar t$, which is a color triplet and a fermion similar to
the quark of the fourth generation.

We have also predicted the existence of the ``b-replaced'' NBS:
$T_s(n_bb-{\mbox{replaced}}) = n_b b + (6t + 6\bar t - n_b t)$ and
$T_f(n_bb-{\mbox{replaced}}) = n_b b + (6t + 5\bar t - n_b t),$
where $n_b, n_{\bar b}=1,2,...6$. The presence of b-quarks in the
NBS leads to the dominance of the isospin singlets.

We have estimated masses of the lightest NBS and showed that the
mass of the scalar T-balls $M_{T_s}$ can be zero, and even
tachyonic: $M_{T_s}^2 < 0$, what leads to the condensation of
T-balls and formation of a new vacuum at the EW-scale.

Also we have estimated masses of the fermionic T-balls predicted
$M_{T_f}\gtrsim 300 \,\,{\mbox{GeV}}$.

It was shown that CDF II Detector experiments searching for heavy
top-like quarks at the Tevatron in $p\bar p$-collisions with
$\sqrt s \backsimeq 1.96$ GeV can observe fermionic $T_f$-balls up
to 500 GeV.

We have considered the processes with T-balls, which can be
observed at LHC, especially the decay $H\to 2T_s$ and the
production of the pair $T_f\ov{T_f}$ combined with the production
of fourth-generation quarks $t'\ov{t'}$-pairs.

We also have constructed the possible form-factors of T-balls.

\section{Acknowledgments}

We deeply thank for the courtesy of CDF collaboration for the
presentation of figures from there.

H.B.N. is grateful to J.~Conway, R.~Erbacher, J.~Frost, H.~Jensen,
C.~Issever, E.~Lytken, K.~Loureiro and A.~Parker for the advices
and fruitful discussions.

L.V.L. thanks A.B.~Kaidalov, O.V.~Pavlovsky and M.A.~Trusov for
the useful discussions.

\end{document}